\documentclass[preprint,prd,tightenlines,nofootinbib,floatfix]{revtex4}
\usepackage{graphicx}
\usepackage{amssymb}
\makeatletter
\DeclareRobustCommand*\cal{\@fontswitch\relax\mathcal}
\makeatother
\def\sket#1{\big|{#1}\big)}

\begin{document}

\title{QCD and Monte Carlo event generators\footnote{\uppercase{P}resented by \uppercase{D}.~\uppercase{E}.~\uppercase{S}oper}}

\author{Zolt\'an Nagy}
\affiliation{
Institute for Theoretical Physics\\
University of Z\"urich\\
Winterthurerstrasse 190\\
CH-8057 Z\"urich, Switzerland\\ 
E-mail: nagyz@physik.unizh.ch
}
\author{Davison~E.~Soper}
\affiliation{
Institute of Theoretical Science\\
University of Oregon\\ 
Eugene, OR 97403, USA\\
E-mail: soper@physics.uoregon.edu
}  

\date{1 July 2006}

\begin{abstract}
Shower Monte Carlo event generators have played an important role in particle physics. Modern experiments would hardly be possible without them. In this talk I discuss how QCD physics is incorporated into the mathematical structure of these programs and I outline recent developments including matching between events with different numbers of hard jets and the inclusion of next-to-leading order effects.
\end{abstract}

\maketitle

\section{A critique of pure perturbation theory}

Before beginning a discussion of shower Monte Carlo event generators, it is well to note that these programs are, for the most part, limited to leading order (LO) perturbative accuracy. On the other hand, there are programs available for a variety of important processes that perform purely perturbative (non-shower) calculations to next-to-leading order (NLO) accuracy. Although these programs have proved to be very useful, it is pertinent to ask why they cannot tell the whole story.

\begin{figure}[htbp]
\vskip - 1.0 cm
\begin{center}
\includegraphics[width = 9 cm]{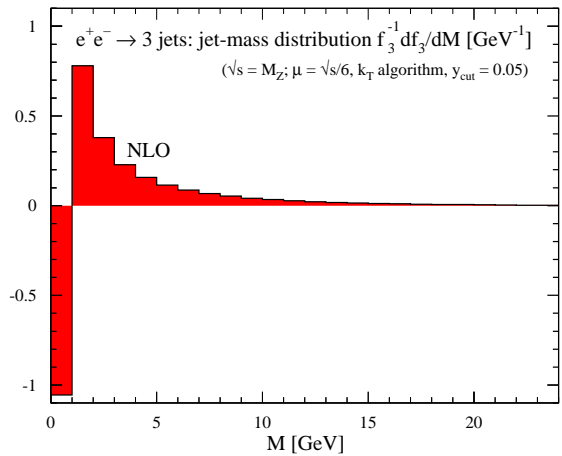}
\vskip - 0.5 cm
\caption{Distribution of the mass $M$ of a jet contributing to the three jet cross section in electron-positron annihilation as determined by a pure NLO calculation\protect{Soper:1998ye}. Note the large negative probability in the first bin.
}
\label{fig:jetmass}
\end{center}
\end{figure}

For this purpose, consider the cross section to produce three jets in electron-positron annihilation (using a suitable definition of what one means by a jet). The ratio of this cross section to the total cross section is the three-jet fraction, $f_3$. Now, $f_3$ is an infrared safe observable that is amenable to calculation at NLO accuracy. In such a calculation, the program produces simulated partonic events with three partons and others with four partons. In either case, if the parton momenta meet certain criteria, the event can be classified as a three jet event. Let us look at this calculation and ask for each jet in each three jet event what the mass of the jet is. Then we can plot the calculated probability of finding a jet in a given bin of jet mass $M$, $f_3^{-1} df_3/dM$. The result of this calculation is shown in Figure~\ref{fig:jetmass}. We see that there are two problems. First, the cross section increases as $M \to 0$. In Figure~\ref{fig:jetmass}, the apparent increase is limited only by the finite bin size. Second, the first bin reflects a large negative cross section. In fact, the cross section for producing a jet with $M=0$ is negative and infinite. Clearly, this calculation is seriously at variance with the real world. Only after we integrate over $M$ do we get a sensible answer.

\begin{figure}[htbp]
\vskip - 1.0 cm
\begin{center}
\includegraphics[width = 9 cm]{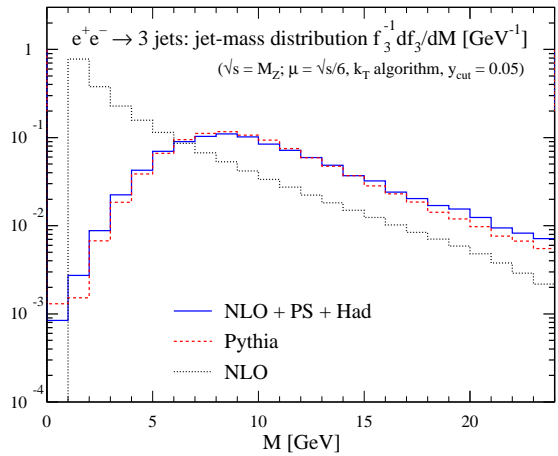}
\vskip - 0.5 cm
\caption{Distribution of the mass $M$ of a jet contributing to the three jet cross section in electron-positron annihilation as determined by an NLO calculation combined with Pythia \protect\cite{MrennaKramerSoper}. The full calculation contains an NLO order hard part followed by parton showering followed by hadronization according to Pythia. This calculation is compared to the corresponding result from Pythia alone and to the pure NLO calculation that was presented in Figure~\ref{fig:jetmass}.
}
\label{fig:jetmass2}
\end{center}
\end{figure}

Now let us look at what a parton shower Monte Carlo program does. In Figure~\ref{fig:jetmass2}, we see that the standard parton shower Monte Carlo program {\tt Pythia} \cite{pythia} produces a sensible result in which the distribution $f_3^{-1} df_3/dM$ peaks at about 8 GeV. This contrasts with the pure NLO prediction. Clearly it would be best to keep the NLO accuracy for $f_3$ while at the same time fixing the internal structure of the jets to be more like what one gets in {\tt Pythia}. This can be done if one keeps track of what the parton shower algorithm does, expands the parton shower effects perturbatively, and subtracts the NLO contribution of the parton shower from the NLO term in the perturbative calculation. The result labeled {\tt NLO\,+\,PS\,+\,Had} combines the NLO calculation with {\tt Pythia} \cite{MrennaKramerSoper}. We see that the result for $f_3^{-1} df_3/dM$ closely follows the pure {\tt Pythia} result. I do not display it as a graph, but the result of this program for just $f_3$ closely follows the pure NLO result.

The program just mentioned is for electron-positron annihilation to make three jets. For hadron-hadron collisions, programs for several important processes are available in the package {\tt MC@NLO}\cite{MCatNLO}.  I will not say more about the technical methods involved in combining ``MC'' with ``NLO,'' but I will comment briefly on one further development later in this talk. In the rest of this talk, I will mainly concentrate on leading order aspects of Monte Carlo event generators.

\section{Showers from the inside out}

\begin{figure}[htbp]
\begin{center}
\includegraphics[width = 4.5 cm]{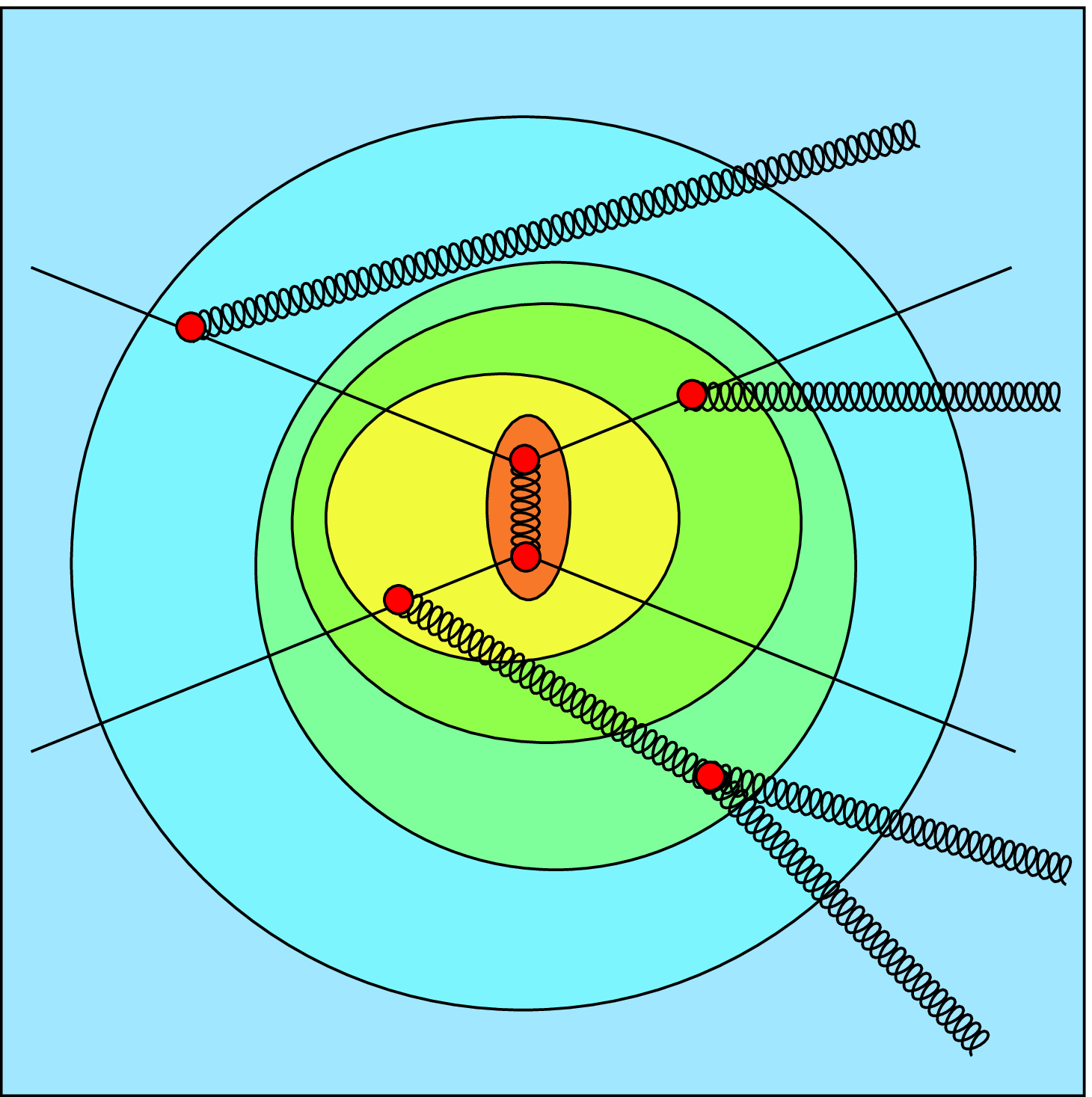}\hskip 1 cm
\includegraphics[width = 4.5 cm]{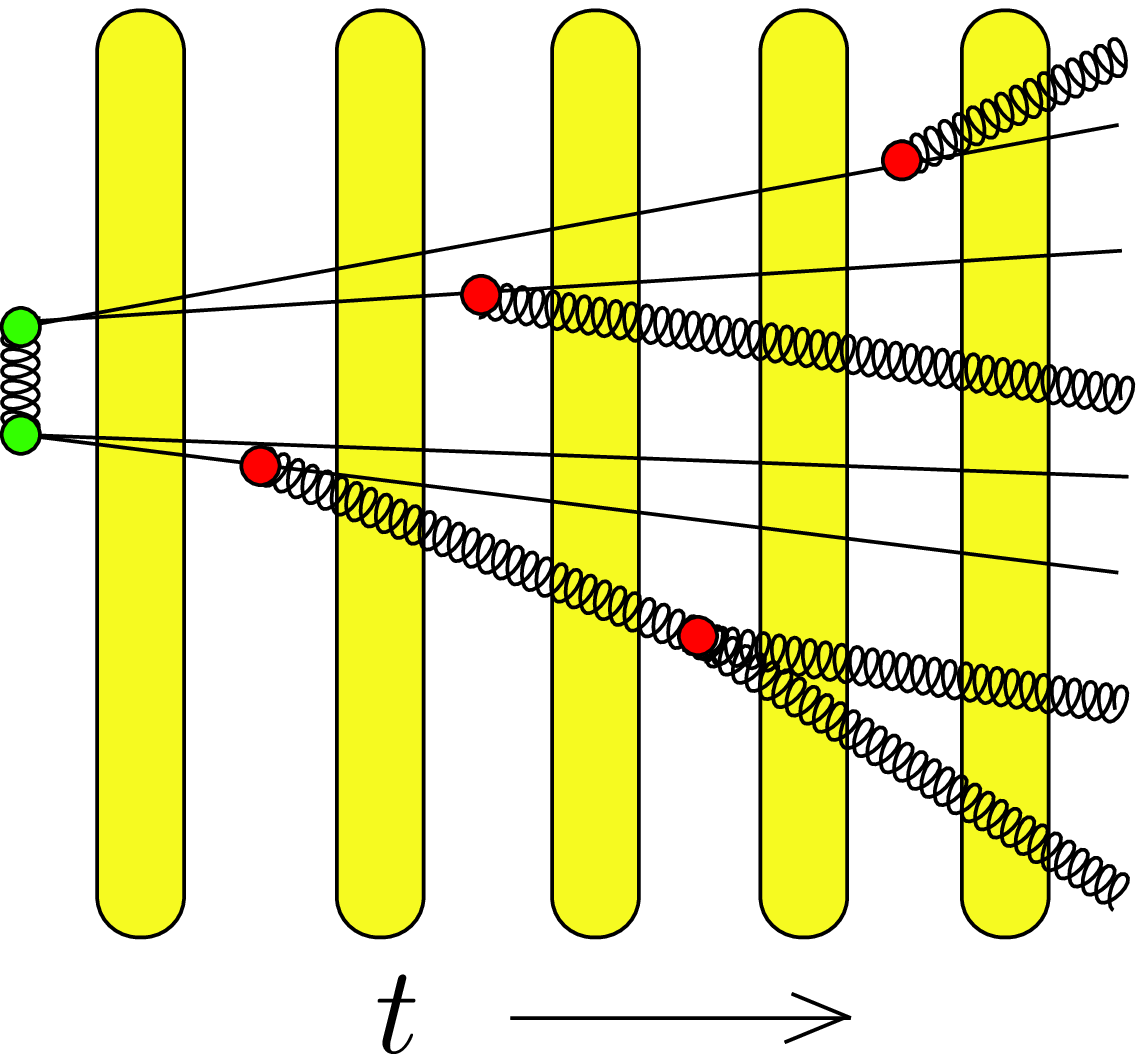}
\caption{The left-hand picture depicts quark-quark scattering, with time proceeding from left to right. The hardest interactions are those toward the center of the picture. These are treated first in a parton shower Monte Carlo program. Thus in Monte Carlo time $t$, we start with the hard process and work toward softer interactions (with some of the particles moving backwards in real physical time), as depicted in the right hand picture. The rounded rectangles represent intervals of Monte Carlo time in which nothing happens.
}
\label{fig:insideout}
\end{center}
\end{figure}

Consider the parton shower picture of hadron-hadron scattering in which there is some sort of hard event, say jet production or squark-pair production. (I leave out the description of the ``underlying event'' that accompanies the hard scattering and I omit entirely the parton shower description of the soft-scattering events that make up most of the hadron-hadron scattering cross section.) The first thing to understand is that the parton shower description starts from the hard scattering and proceeds toward softer scatterings. For final state partons, one is thus working forwards in time, but for the initial state partons one is working backwards in time. This is depicted in Figure \ref{fig:insideout}. In the right hand part of the figure, $t$ is Monte Carlo time, for which $t = 0$ represents the hardest interaction and $t \to \infty$ as parton virtualities tend to zero. Thus some of the lines in the right hand picture are moving backward in physical time. Although the development of the parton shower description of hadron scattering dates from about 1980 \cite{Odorico:1980gg}, it was not until somewhat later that this backwards evolution scheme was developed \cite{backward}.

I should mention that in the description that I present here, shower evolution is organized according to a hardness measure like virtuality or the transverse momentum in a splitting. The important program {\tt Herwig} \cite{herwig} is organized differently, with splittings at the widest angles done first. I do not discuss the {\tt Herwig} scheme here. Rather, the description here is more attuned to the program {\tt Pythia} \cite{pythia}.

\section{Color coherence}

\begin{figure}[htbp]
\begin{center}
\includegraphics[width = 3.5 cm]{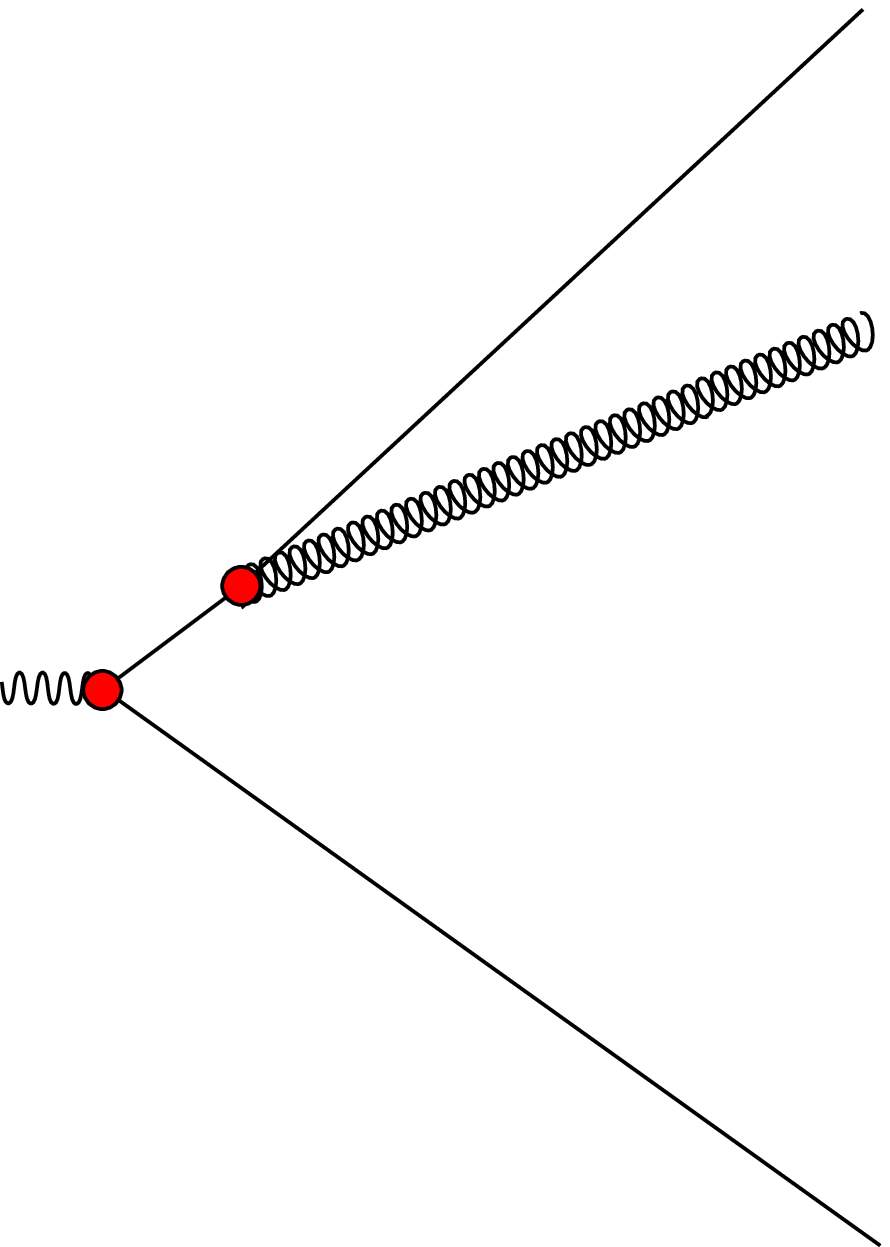}\hskip 1 cm
\includegraphics[width = 3.3 cm]{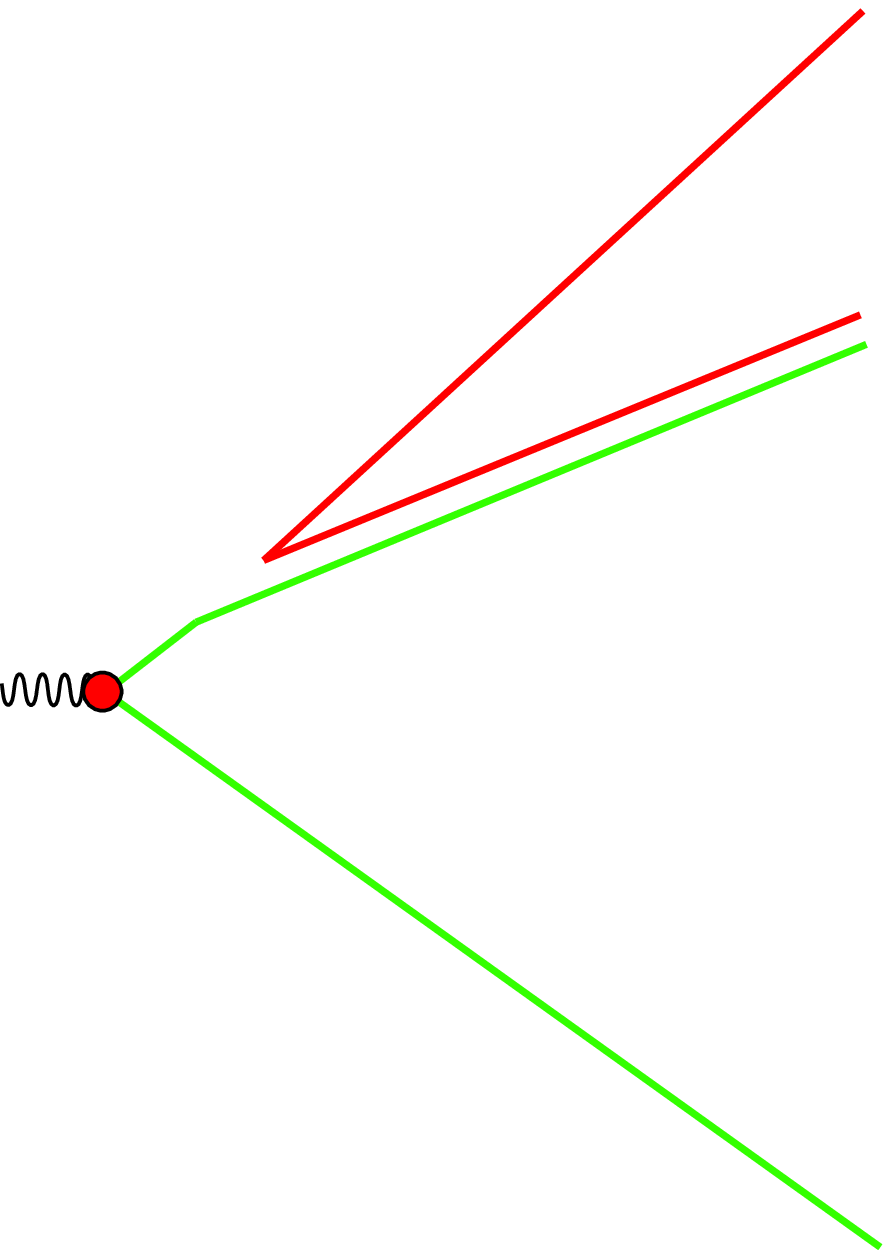}
\caption{ 
The left-hand picture depicts the production of three hard jets in an electron-positron collision. The right-hand picture depicts the color structure of the final state. To leading order in $1/N_c^2$, the gluon can be treated as a color $\bf 3$ state moving in the same direction as a color $\bf \overline 3$ state. In this approximation the final state has two independent $\bf 3\, \bar 3$ color dipoles.
}
\label{fig:coherence}
\end{center}
\end{figure}

Now let us think about soft gluon radiation. I consider three jet production in electron-positron annihilation as an example. At the Born level, one has a $q\, \bar q\,g$ final state. The gluon is a color $\bf 8$, but to leading order in an expansion in powers of $1/N_c^2$, where $N_c = 3$ is the number of colors, the gluon can be considered to be a $\bf 3\, \bar 3$ state. Then the outgoing quark and the $\bf \bar 3$ part of the gluon constitute a $\bf 3\, \bar 3$ dipole, while the outgoing antiquark and the $\bf 3$ part of the gluon constitute another $\bf 3\, \bar 3$ dipole, as depicted in Figure~\ref{fig:coherence}. (These are ``dipoles'' in the sense that there is a color charge and an opposite color charge moving away from each other. However, the color charge separation is not small as in the traditional multipole approximation.)

\begin{figure}[htbp]
\begin{center}
\includegraphics[width = 5 cm]{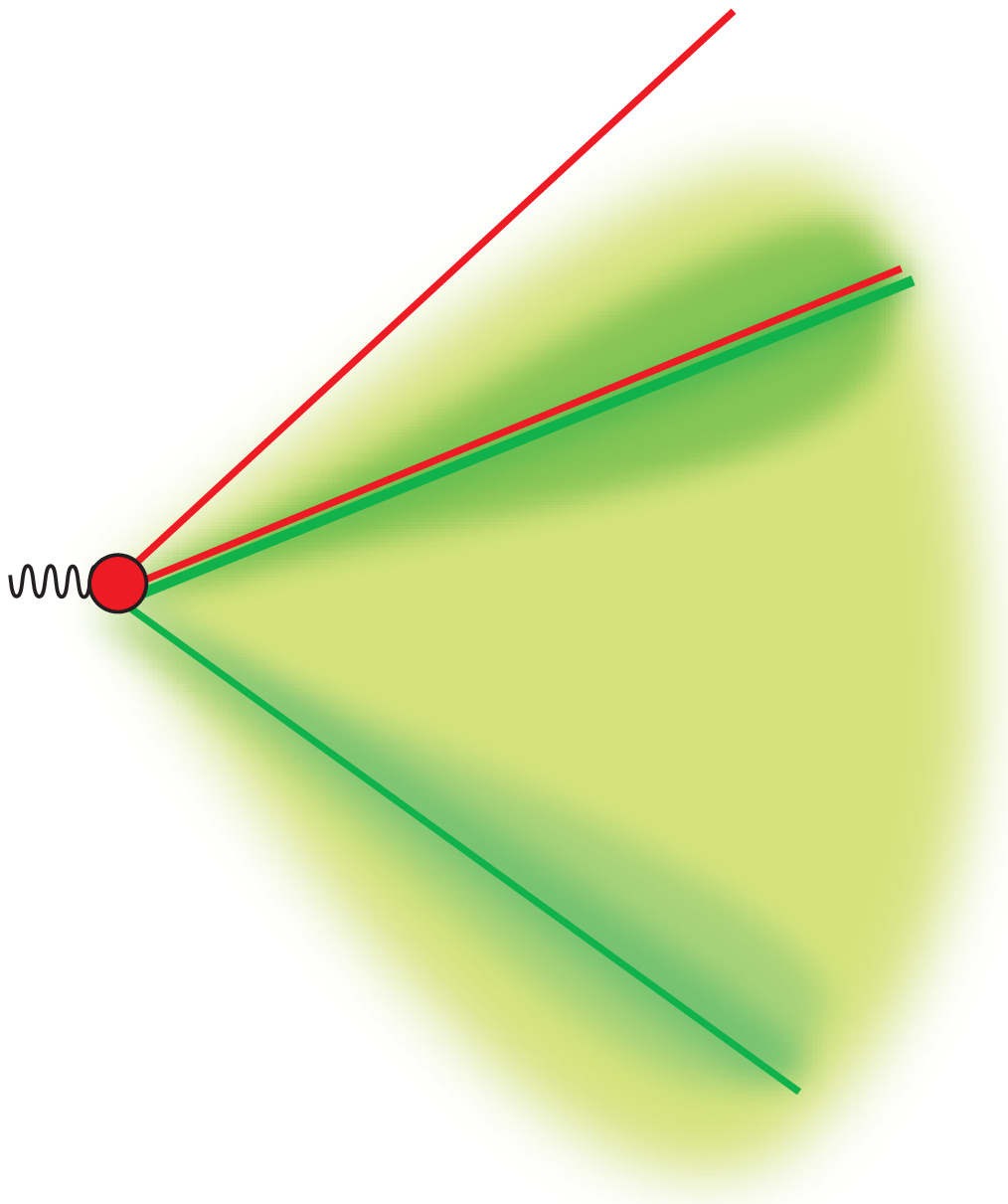}\hskip 1 cm 
\raisebox{0.8 cm}{
\includegraphics[width = 4.6 cm]{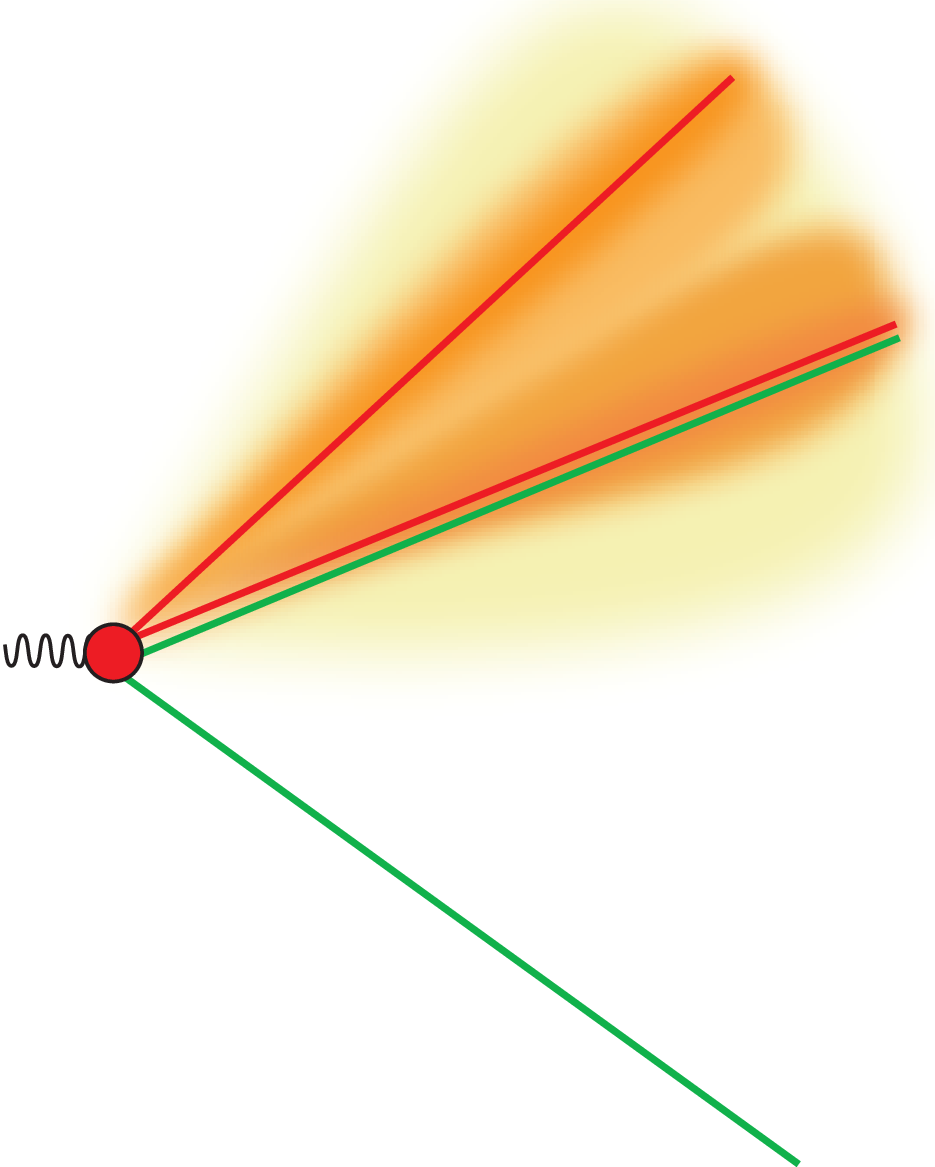}
}
\caption{ 
Radiation from the two dipoles of Figure~\ref{fig:coherence}. For each dipole, the radiation is concentrated along the direction of the two outgoing partons and also contains a wider angle component that is spread over an angular range that is roughly that subtended by the two parton momenta.
}
\label{fig:dipoleradiation}
\end{center}
\end{figure}

The two dipoles will radiate soft gluons. Given the (approximate) color structure, the two dipoles radiate independently: there is no quantum interference between a gluon radiation from one dipole and radiation from the other dipole. The radiation pattern is depicted in Figure~\ref{fig:dipoleradiation}. For each dipole, there is soft-collinear radiation that is concentrated in the directions of the two outgoing partons for that dipole. There is also a wider angle component that is, approximately, spread over the angular region between the parton directions. Thus the wide angle dipole has soft radiation spread over a wide angular region while the narrow angle dipole has soft radiation spread over a narrow angular region.

In a parton shower Monte Carlo program, one can work to leading order in $1/N_c^2$ (as parton shower programs generally do) and make sure that the parton splitting formulas properly take into account the interference between gluons emitted from the two parts of a color dipole. The program {\tt Ariadne} is based on this kind of picture \cite{ariadne}. The latest version of {\tt Pythia} is also based on a dipole picture \cite{pythia,SjostrandSkands}. The present authors have found that the Catani-Seymour dipole formalism \cite{CataniSeymour} for generating the subtractions for perturbative NLO calculations is also quite useful as the basis for splittings in a parton shower \cite{NagySoperJHEP,NagySoperRingberg}.

There is another way to do this. One can simply generate independent emissions from each parton and then impose a restriction on the angles of the emissions. This is the method of {\tt Herwig} \cite{herwig}. In {\tt Herwig}, a wide angle soft gluon emission as depicted in the left-hand part of Figure~\ref{fig:dipoleradiation} is generated first, before the splitting of the quark into a hard quark and a hard gluon. The algorithm enforces that the angles between daughter partons in a splitting decrease for splittings generated later in the algorithm evolution. The recognition of the importance of this ordering was important in the development of parton shower algorithms \cite{Webber}.

\section{Shower evolution in pictures}

Shower evolution can be represented using an evolution equation of the form
\begin{equation}
U(t_3,t_1) = N(t_3,t_1)
+ \int_{t_1}^{t_3}\! dt_2\ 
U(t_3,t_2)\,{\cal H}(t_2)\,N(t_2,t_1)
\;\;.
\label{eq:evolution}
\end{equation}
Here $U(t_3,t_1)$ is a linear operator acting on a function $\sket{{\cal A}(t_1)}$ giving the probability for having a given partonic state at evolution time $t_1$. Then $U(t_3,t_1)\sket{{\cal A}(t_1)}$ is the probability function for having a particular partonic state at a later evolution time $t_3$. The evolution equation expresses $U$ as a sum of two terms. In the first term, the partonic state stays the same. This is represented as a no splitting operator $N(t_3,t_1)$ that inserts a Sudakov factor giving the probability that no splitting occurred between times $t_1$ and $t_3$. This factor is an exponential of minus the integral from $t_1$ to $t_3$ of the differential splitting probability. In the second term, there is a no splitting operator $N(t_2,t_1)$, followed by a splitting operator ${\cal H}(t_2)$ at time $t_2$, followed by complete evolution $U(t_3,t_2)$ for times after the splitting. There is an integration over the intermediate time $t_2$ at which the splitting occurs. The $1 \to 2$ splitting operator ${\cal H}(t_2)$ uses an approximation to the QCD squared matrix element.

\begin{figure}[htbp]
\begin{center}
\includegraphics[width = 6 cm]{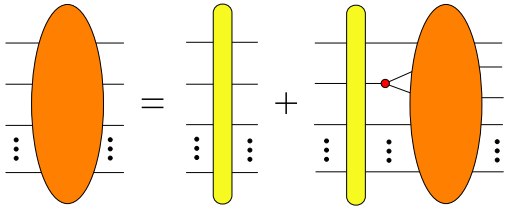}
\caption{ 
The evolution equation in pictures, as described in the text.}
\label{fig:evolution}
\end{center}
\end{figure}

This shower evolution can be represented graphically as in Figure~\ref{fig:evolution}. The ovals represent the complete shower evolution operator $U$.   The narrow rounded rectangles represent the Sudakov operator $N$. The splitting is represented by a small circle with one parton coming in and two partons going out. 

\begin{figure}[htbp]
\begin{center}
\includegraphics[width = 10 cm]{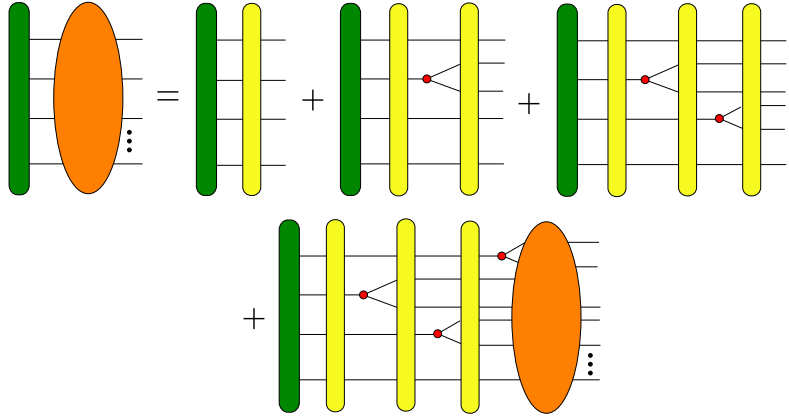}
\caption{ 
Calculation of a shower starting with a $2 \to 2$ hard cross section (dark rounded rectangle). The shower evolution operator has been iterated twice, so that the first term represents no splitting, the second term has one splitting, the third term has two splittings, and the final term contains contributions with three or more splittings.
}
\label{fig:iterated}
\end{center}
\end{figure}

To generate a cross section with a shower Monte Carlo event generator, one can start with a hard squared matrix element for $2 \to 2$ scattering, then apply the shower operator to the two incoming and two outgoing partons. When the shower evolution equation is iterated, one obtains terms representing $n = 0,1,2,\dots$ splittings with Sudakov factors for the intervals with no splittings, as depicted in Figure~\ref{fig:iterated}.

\section{An improved shower}

\begin{figure}[htbp]
\begin{center}
\includegraphics[width = 3.6 cm]{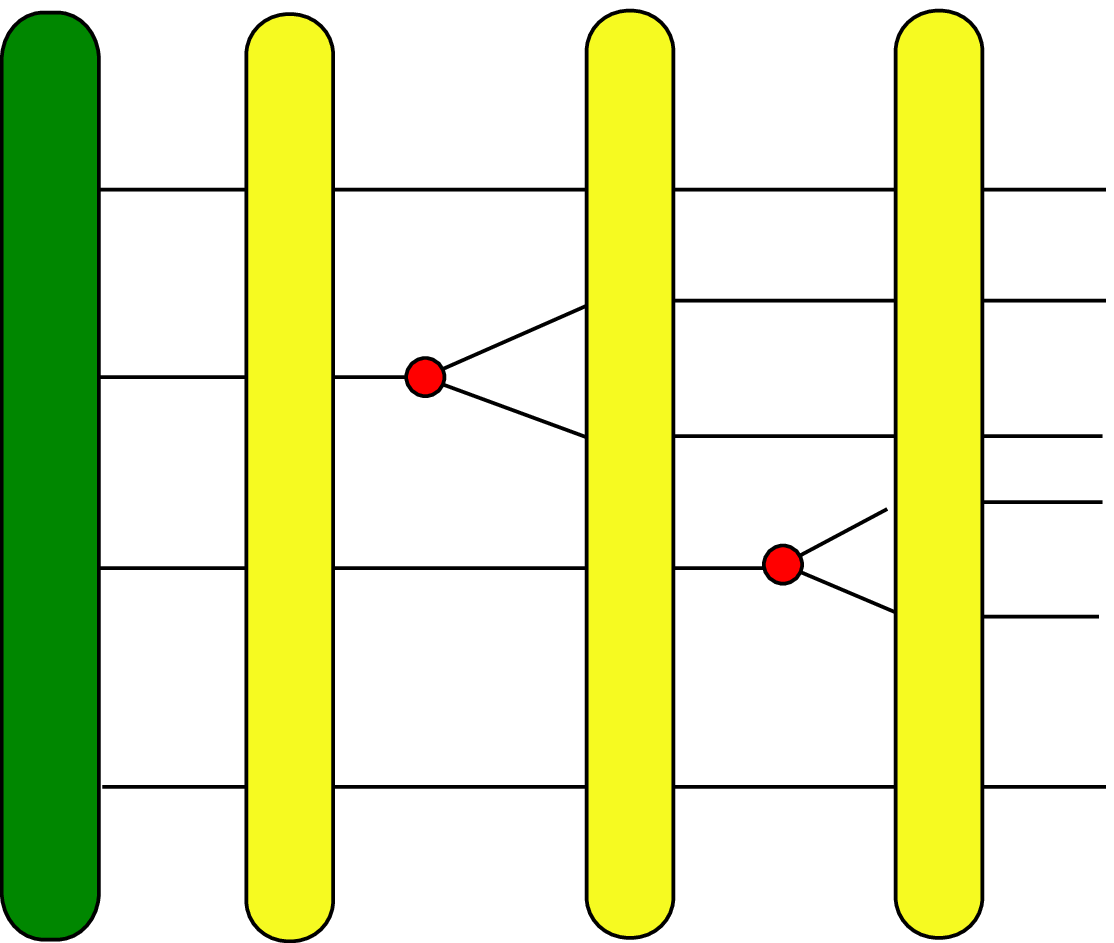}\hskip 2 cm
\includegraphics[width = 3.0 cm]{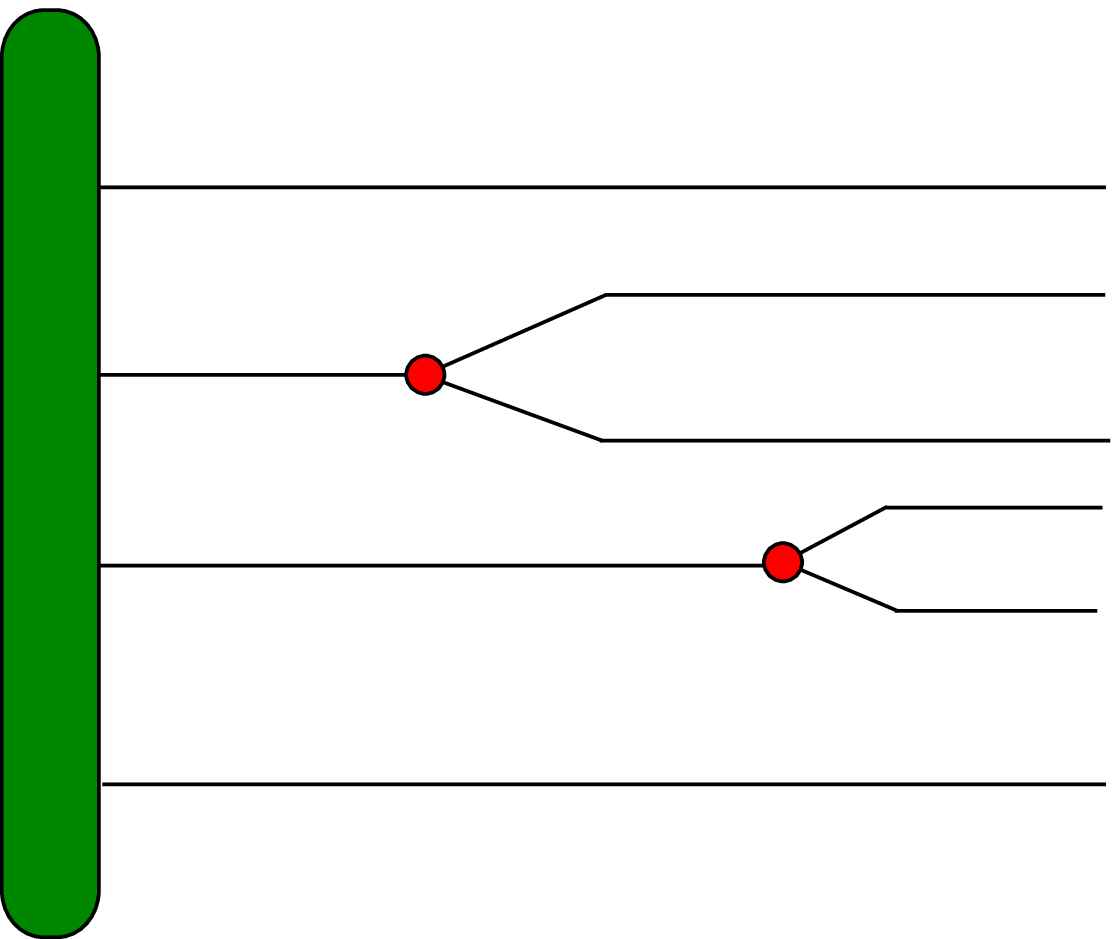}\hskip 2.5 cm
\includegraphics[width = 0.36 cm]{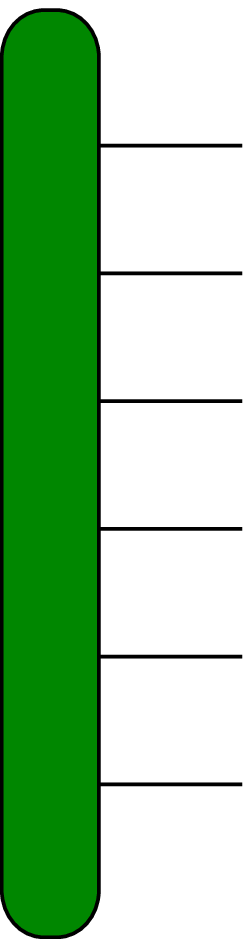}
\caption{ 
The left-hand picture is the $2 \to 4$ cross section in shower approximation. The center picture is the shower approximation omitting the Sudakov factors. The right hand picture is the exact tree level $2 \to 4$ cross section. The cross section based on splitting functions (middle picture) is a collinear/soft approximation to this.
}
\label{fig:deficiency}
\end{center}
\end{figure}

The standard shower depicted in Figure~\ref{fig:iterated} has a deficiency, which is illustrated in Figure~\ref{fig:deficiency}. The left-hand picture depicts a term contributing to the standard shower. In this term, there are Sudakov factors and $1 \to 2$ parton splitting functions. If we omit the Sudakov factors, we have the $1 \to 2$ parton splittings as depicted in the middle picture. These splittings are approximations based on the splitting angles being small or one of the daughter partons having small momentum. Thus the shower splitting probability with two splittings approximates the exact squared matrix element for $2 \to 4$ scattering. The approximation is good in parts of the final state phase space, but not in all of it. Thus one might want to replace the approximate squared matrix element of the middle picture with the exact squared matrix element of the right-hand picture. However, if we use the exact squared matrix element, we lack the Sudakov factors.

\begin{figure}[htbp]
\begin{center}
\includegraphics[width = 10 cm]{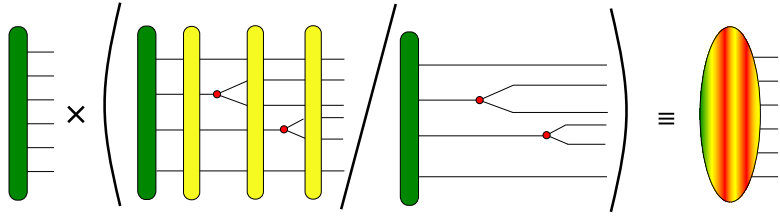}
\caption{ 
An improved version of the $2 \to 4$ cross section. We take the shower approximation, divide by the approximate collinear squared matrix element, and multiply by the exact tree level squared matrix element. The graphical symbol on the right hand side represents this Sudakov reweighted cross section.
}
\label{fig:improved}
\end{center}
\end{figure}

One can improve the approximation as illustrated in Figure~\ref{fig:improved}. We reweight the exact squared matrix element by the ratio of the shower approximation with Sudakov factors to the shower approximation without Sudakov factors. The idea is to insert the Sudakov factors into the exact squared matrix element. This is the essential idea in the paper of Catani, Krauss, Kuhn, and Webber \cite{CKKW}. They use the $k_T$ jet algorithm to define the ratio needed to calculate the Sudakov reweighting factor.

There is a further step in implementing this idea. CKKW divide the shower evolution into two stages, $0 < t < t_{\rm ini}$ and $t_{\rm ini} < t < t_{\rm f}$, where $t_{\rm ini}$ is a parameter that represents a moderate $P_T$ scale and $t_{\rm f}$ represents the very small $P_T$ scale at which showers stop and hadronization is simulated.

\begin{figure}[htbp]
\begin{center}
\includegraphics[width = 10 cm]{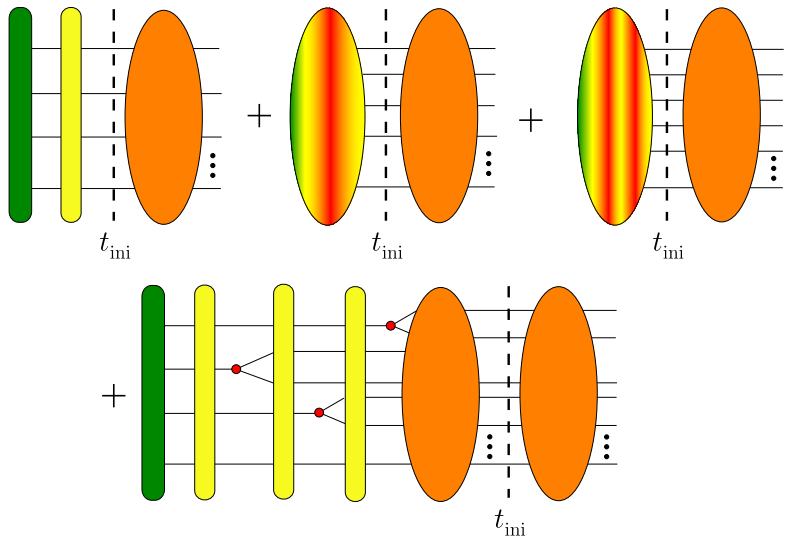}
\caption{ 
Shower with CKKW jet number matching. The calculation for $n$ jets at scale $t_{\rm ini}$ is based on the Sudakov reweighted tree level cross section for the production of $n$ partons.
}
\label{fig:CKKW}
\end{center}
\end{figure}

With this division, the Sudakov reweighting can be performed for the part of the shower at scale harder than $t_{\rm ini}$, as depicted in Figure~\ref{fig:CKKW}. The first term has no splittings at scale harder than $t_{\rm ini}$. In the second term there is one splitting, generated via the exact matrix element with a Sudakov correction as discussed above. In the next term there are two splittings. If we suppose that we do not have exact matrix elements for more than $2 \to 4$ partons, states at scale $t_{\rm ini}$ with more partons are generated with the ordinary parton shower. However, this contribution is suppressed by factors of $\alpha_s$. Evolution from $t_{\rm ini}$ to $t_{\rm f}$ is done via the ordinary shower algorithm.

To state the main idea of this jet number matching in a little different language, we can consider the cross section for an observable $F$. In the CKKW method, we break $\sigma[F]$ into a sum of contributions $\sigma_m[F]$ from final states with $m$ jets at resolution scale $t_{\rm ini}$. Then $\sigma_m[F]$ is evaluated using the exact tree level matrix element for $2 \to m$ parton scattering, supplemented by Sudakov reweighting and further supplemented by showering of the $m+2$ partons at scales softer than $t_{\rm ini}$. If $F$ is an infrared safe observable, this method gets $\sigma_m[F]$ correct to the leading perturbative order, $\alpha_s^m$. The method can be extended. The present authors have shown (at least for the case of electron-positron annihilation) how to get $\sigma_m[F]$ for an infrared safe observable correct to next-to-leading order, $\alpha_s^{m+1}$ \cite{NagySoperJHEP}. The required NLO adjustments are a little complicated, so I do not discuss them here.

\section{An alternative shower improvement}

\begin{figure}[htbp]
\begin{center}
\includegraphics[width = 9 cm]{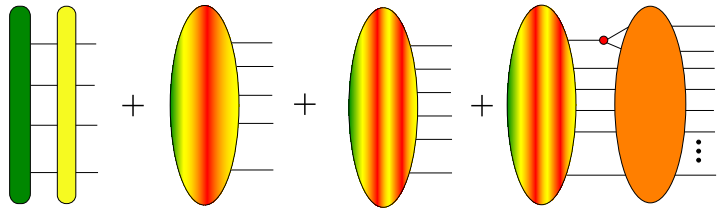}
\caption{ 
Alternative shower improvement based on using the exact tree level matrix elements to produce a given number $N$ of partons rather than producing an indefinite number of partons at a given scale $t_{\rm ini}$.
}
\label{fig:CKKWalt}
\end{center}
\end{figure}

There is an alternative way to organize the shower improvement so as to include exact tree level matrix elements \cite{NagySoperRingberg}. One does not really need to split the evolution at a scale $t_{\rm ini}$. Suppose that one has the exact tree level matrix elements for $2 \to n$ partons for $n \le N$. Then the partonic cross section at a final very soft scale $t_{\rm f}$ before hadronization is the sum of the $2 \to 2$, $2 \to 3$, \dots $2 \to N$ cross sections with Sudakov factors plus one more term, which is the most important term. In the last term, we have the Sudakov improved $2 \to N$ squared matrix element in which the softest splitting has scale $t$ and we integrate over $t$. This is convoluted with the simple shower approximation for splittings softer than $t$, down all the way to $t_{\rm f}$. This is depicted in Figure~\ref{fig:CKKWalt}. The terms before the last one are included in the calculation but are not important because they contain the Sudakov suppression for only a small number of splittings to occur down to a very soft scale $t_{\rm f}$. In the term that really matters, we use the Sudakov improved $2 \to N$ squared matrix element with an arbitrary number of further splittings generated in the collinear/soft approximation, all of this with Sudakov suppression factors.

\section{Outlook}

Parton shower event generators have proved to be an essential tool for particle physics. These tools have been undergoing substantial improvement in recent years. There are quite a number of very good people working on this subject and some of the most recent progress will be reported at this conference.

\section*{Acknowledgments}
This work was supported in part the United States Department of Energy
and by the Swiss National Science Foundation (SNF) through grant no.\
200020-109162 and by the Hungarian Scientific Research Fund grants OTKA
K-60432.


-------------------------------------------------------------------------------
\end{document}